\documentclass[a4paper,11pt]{article}
\usepackage{graphicx}

\begin{document}

\title{POSSIBLY EXACT SOLUTION FOR THE MULTICRITICAL POINT
OF FINITE-DIMENSIONAL SPIN GLASSES}
\author{Hidetoshi Nishimori\footnote{e-mail: nishimori@phys.titech.ac.jp},
 ~Koujin Takeda\footnote{e-mail: takeda@stat.phys.titech.ac.jp}
 ~and Tomohiro Sasamoto\footnote{e-mail: sasamoto@stat.phys.titech.ac.jp} \\
\\
Department of Physics, Tokyo Institute of Technology, \\
Oh-okayama, Meguro-ku, Tokyo 152-8551, Japan}

\maketitle
\begin{abstract}
After briefly describing the present status of the spin glass theory,
we present a conjecture on the exact location of the multicritical
point in the phase diagram of finite-dimensional spin glasses.
The theory enables us to understand in a unified way many
numerical results for two-, three- and four-dimensional models including
the $\pm J$ Ising model, random Potts model, random lattice gauge theory,
and random $Z_q$ model.
It is also suggested from the same theoretical framework
that models with symmetric distribution of
randomness in exchange interaction have no finite-temperature transition
on the square lattice.
\end{abstract}

\section{Introduction}
Spin glasses are magnetic materials in which spins are randomly frozen.
Theoretical studies of spin glasses started with the paper by
Edwards and Anderson in 1975\cite{EA} in which they proposed
to model the problem in terms of the following Hamiltonian,
now termed the Edwards-Anderson model,
\begin{equation}
  H=-\sum_{\langle i,j\rangle} J_{ij}S_i S_j.
\label{EA}
\end{equation}
In the present contribution we will mainly discuss the Ising model ($S_i=\pm 1$)
with binary ($J_{ij}=\pm J$) or Gaussian quenched randomness
in exchange interactions.
The infinite range version of this model was proposed and analyzed by Sherrington
and Kirkpatrick soon afterwards using the replica method\cite{SK} .
These two papers set the landmark of the spin glass theory,
and a number of facts are now known, and yet many problems still remain to be solved,
particularly in finite dimensions.

Two central issues concerning the static properties of spin glasses are, first,
whether or not a spin glass phase exists under a given condition and, second,
what characteristics the spin glass phase has if it exists.  
A successful mean-field theory has been constructed with
replica symmetry breaking\cite{Parisi} that solved both of these problems.
A spin glass phase, in which spins are randomly frozen, does exist at low temperatures
for $J_0$ (the center of distribution of Gaussian
exchange interaction values) close to 0.
Ferromagnetic and paramagnetic phases also exist in appropriate parameter regions.
In the mean-field model
the spin glass phase shows peculiar behavior that the free energy has
infinitely many minima, {\it i.e.} infinitely many stable states, quite
differently from the usual ferromagnet which has essentially a unique minimum.

\begin{figure}
\begin{center}
 \includegraphics[width=6cm]{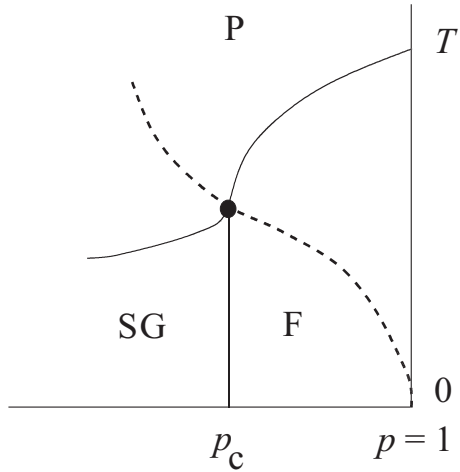}
 \label{fig:phase-diagram-finite-d}
\caption{A typical phase diagram of the finite-dimensional $\pm J$ Ising model, in which
the interaction is ferromagnetic with probability $p$ and antiferromagnetic with $1-p$.
The multicritical point is marked as a black circle.
The Nishimori line is shown dashed.}
\end{center}
\end{figure}

It is important to investigate whether or not these mean-field predictions
apply to realistic finite-dimensional systems.  Efforts toward this direction
have so far clarified that a spin glass phase is very likely to exist in three 
and higher dimensions
for the random Ising system as depicted in Fig. \ref{fig:phase-diagram-finite-d}.
Still under debate are the nature of the spin glass phase, whether or not it has a very
complex mean-field-like structure, and how Heisenberg-like isotropic spin systems
show spin glass behavior in experiments.

A serious problem in the research activities on finite-dimensional cases, which
is the main topic of the present contribution, is that mathematical analysis is
extremely difficult and most of the established (and still debated) facts come
from numerical studies.  This is in marked contrast with the mean-field theory,
where a very sophisticated scheme of replica symmetry breaking and related
methods serve as powerful analytical tools.

We were therefore surprised to have found that the well-known technique of duality
transformation, in conjunction with arguments using gauge symmetry, enables us to
predict the exact (correctly speaking, possibly exact) location of the multicritical
point, where the ferromagnetic phase ceases to exist in the phase diagram (see Fig. 1),
not only in the two-dimensional $\pm J$ Ising spin glass but also in 
other systems in two, three, and four dimensions.
Our line of reasoning is not completely rigours mathematically, and the results
remain to be a conjecture at this moment.  We nevertheless believe that
a formal proof will come before too long, in which case the present results
may serve as an important step toward a comprehensive analytical solution of
the spin glass problem in finite dimensions.
The following sections describe our results and a summary of the steps to reach them.

\section{Phase Diagram in Finite Dimensions: Known Facts}

A typical phase diagram is drawn in Fig. \ref{fig:phase-diagram-finite-d}.
The central topic in the present contribution is the identification of the exact
location of the multicritical point marked by a black circle.
Before presenting the results, let us summarize what has already been known
about the phase diagram of finite-dimensional spin glasses, in particular
the $\pm J$ Ising model, mainly by symmetry arguments\cite{HN2001}.

An important fact is that the exact value of the energy,
averaged over quenched randomness, can be evaluated with the result
$E=-N_BJ\tanh K$,
where $N_B$ is the number of bonds in the system and $K=\beta J$.
This formula applies as long as the temperature and probability are
related as $e^{-2K}=(1-p)/p$,
which defines a line in the $T$-$p$ phase diagram, called the Nishimori line (NL)
shown dashed in Fig.  \ref{fig:phase-diagram-finite-d}.
A remarkable aspect is that the exact expression given above is valid for any
system with any range of interaction on any lattice, which is shared
by all the following results in this section.
It should be noted that the energy $-N_BJ\tanh K$ has no singularity although
the NL clearly crosses the boundary between ferromagnetic and paramagnetic
phases; indeed it is believed that the crossing is at the multicritical point.
Thus we can focus our attention to the NL to solve the problem of identification of
the exact location of the multicritical point.
Only the energy is non-singular along the NL, and other physical quantities including the
free energy, specific heat and magnetic susceptibility are of course singular
at the multicritical point even along the NL.

It can also be shown that the ferromagnetic order parameter $m$ is exactly equal
to the spin glass order parameter $q$ on the NL ($m=q$) under quite general conditions.
Since the spin glass phase has finite $q>0$ and vanishing $m=0$, it immediately follows
that the spin glass phase, if it exists, cannot have the NL in it, and is most
likely to lie below it as displayed in Fig. \ref{fig:phase-diagram-finite-d}.
This is also remarkable because this rigorous constraint has been shown to hold for the
possible region where such a phase is allowed to exist when we do not know for sure by
analytical methods
if a spin glass phase exists in finite dimensions.

It has been proved that there is no replica symmetry breaking on the NL
in the sense that the distribution function of the spin glass order parameter is
equal to that of the ferromagnetic counterpart, $P(q)=P(m)$, the latter 
distribution function being of quite simple trivial structure. 
Thus the mean-field-like complex phase with non-trivial $P(q)$ can exist, if at all,
only away from the NL, again most plausibly below it.

These and many other facts, derived in a mathematically rigorous way without
recourse to the replica method but using only gauge symmetry of the system,
constitute the almost only set of exact/rigorous results for finite-dimensional
spin glass models\cite{HN2001}.
The outstanding generality of the theory, {\it i.e.} its applicability to
any lattice, any dimension or any range of interaction, is its
strength and, simultaneously, its weakness.  The reason for weakness is that
the theory is unable to predict lattice-specific properties such as the
location of the multicritical point.

Developments in the last few years\cite{NN,MNN,TN,TSN} have completely changed
the situation, and we are now able to predict the exact location of the
multicritical point for self-dual lattices in finite dimensions.
For non-self-dual lattice pairs such as the triangular and hexagonal lattices,
our duality argument allows us to relate the locations of
multicritical points of two mutually dual systems in a compact formula.
We will also develop an argument that the spin glass with symmetric randomness distribution
has no finite temperature transition on the square lattice.

\section{Results}

As has been mentioned, our results have the status of conjecture,
not rigorously proved facts.  To convince the reader that our conjecture
is most likely to be correct, it should be helpful to show the final
values that are compared quite favorably with numerical simulation
results as in Table 1.

\vspace{5mm}
\begin{table}[h]
\begin{center}
\begin{tabular}{@{}lll@{}}
\hline
  Model & Typical numerical result   & Our prediction \\
\hline
  SQ Ising                  & 0.8894(9)\cite{IO}          & 0.889972 \\
  SQ Gaussian               & 1.00(2)\cite{OzekiG}        & 1.021777 \\
  SQ 3-Potts                & 0.921(1)\cite{Pico}         & 0.920269 \\
  $4d$ lattice gauge        & 0.890(2)\cite{Ichinose}       & 0.889972 \\
  TR                        & 0.835(5)$(=p_{c1})$\cite{TSN}          & --- \\
  HEX                       & 0.930(5)$(=p_{c2})$\cite{TSN}          & --- \\
  TR + HEX                  & $H(p_{c1})+H(p_{c2})=1.01(3)$ & $H(p_{c1})+H(p_{c2})=1$\\
  $3d$ Ising (RBIM)         & 0.7673(3)$(=p_{c1})$\cite{OI}          & --- \\
  $3d$ gauge (RPGM)         & 0.967(4)$(=p_{c2})$\cite{OAIM-WHP}     & --- \\
  RBIM+RPGM                 & $H(p_{c1})+H(p_{c2})=0.99(2)$ & $H(p_{c1})+H(p_{c2})=1$\\
\hline
\end{tabular}
\caption{Location of multicritical point by numerical studies and our prediction.
SQ stands for the square lattice, and TR/HEX for the triangular/hexagonal lattices,
respectively. The values are for $p_c$ of the $\pm J$ model
except for the Gaussian randomness for which the value is for $J_{0c}/J$.
Spin variables are Ising excepting the three-state Potts model as indicated.
Our analysis gives definite values for self-dual systems for which explicit
numbers are given in the third column whereas, for mutually dual pairs such as the
triangular and hexagonal lattices, the prediction is that the two values are related
in a simple formula $H(p_{c1})+H(p_{c2})=1$, where $H(p)$ is the binary
entropy $-p\log_2 p-(1-p)\log_2 (1-p)$.}
\end{center}
\end{table}

A remarkable fact to bear in mind is that the numbers given in the second column
of Table 1 have come out of independent numerical investigations whereas
the third column has been filled in by a single unified theoretical framework,
an essence of which is to be explained shortly.
We therefore have strong confidence that the agreement observed in this
Table between numerics and analysis is not due to accidental coincidence.

The following are the explicit formulas for the above prediction. For self-dual lattices,
the critical concentration $p_c$ for the $\pm J$ Ising model satisfies
$2H(p_{c})=1$ irrespective of the lattice structure.
This equation is solved to give $p_c=0.889972$.
If the lattice is not self dual,
the pair of mutually dual lattices are related by the formula
$H(p_{c1})+H(p_{c2})=1.$
In the Gaussian case on self-dual lattices, the critical value of $J_{0}$ satisfies
\begin{equation}
 2\int_{-\infty}^{\infty} du\, P(u)\log_2 (1+e^{-2J_0 u/J^2})=1,
\end{equation}
where $P(u)$ is the Gaussian distribution.
The formula for the $q$-state Potts model is
\begin{equation}
  2 \{-(1-(q-1)p_c) \log_q (1-(q-1)p_c)-(q-1)p_c \log_q p_c \} =1.
\end{equation}
All of these results have been derived using duality tranformation combined with
gauge symmetry.

\section{Steps to the Results: Duality and Gauge Symmetry}

Let us next explain our basic ideas for the Ising system on a self-dual lattice.
 Details are found in Refs. \cite{MNN,TN,TSN}.
A preliminary report based on a more naive argument,
but leading to the same conclusion, appeared in Ref. \cite{NN}.

The well-known method to identify the critical point of the ferromagnetic
Ising model on a self-dual lattice uses invariance of the partition function
under duality transformation of the temperature, $Z(T)=Z(T^*(T))$,
where $T^*(T)$ is the dual temperature of the original $T$ and is a
monotone decreasing function of $T$.  If the critical point is unique,
it should be shared by both sides of $Z(T)=Z(T^*(T))$, and hence
the fixed-point condition of duality transformation $T=T^*(T)$
yields the location of the critical point.
Since the partition function can also be regarded as a function of
the local Boltzmann factors $x_0(=e^K)$ (for parallel spin pair)
and $x_1(=e^{-K})$ (for antiparallel pair), the above duality expression
is also written as
\begin{equation}
  Z(x_0, x_1)=Z(x_0^*,x_1^*).
 \label{dual-ferro2}
\end{equation}
The fixed-point condition $T=T^*(T)$ is equivalent to $x_0=x_0^*$ as well as
to  $x_1=x_1^*$.

A straightforward application of the same idea as above to systems with
quenched randomness with positive and negative interactions leads to
imaginary couplings in the dual system.  To avoid such a difficulty,
it is convenient to introduce replicas and average the partition function,
replicated $n$ times, over the randomness, which yields the translationally
invariant effective partition function $Z_n$.  The goal is to analyze the
system properties in the limit $n\to 0$, but we consider integer-$n$ cases
for the moment.
The replicated partition function $Z_n$ is completely specified by giving the
values of local Boltzmann factors because the system is translationally invariant
after configurational average.
Since each site has $n$ spins, the number of values of local Boltzmann factors
is $n+1$, which we denote as $x_0, x_1, \cdots , x_n$.  The leading factor
$x_0$ denotes the Boltzmann weight for all-parallel spin configuration between
neighboring sites (see Fig. 2), $x_1$ is for the case with single
antiparallel pair with the remaining being parallel, and so on.
\begin{figure}
\begin{center}
 \includegraphics[width=7cm]{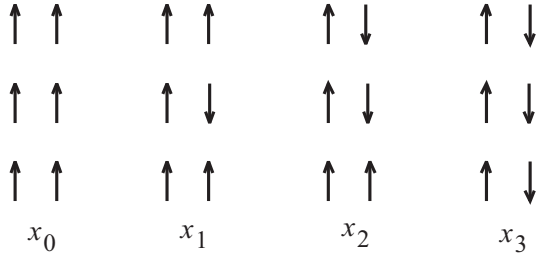}
 \label{fig:BF}
\caption{Possible relative spin configurations between neighboring sites
for the $n=3$ replicated system.  $x_k$ stands for the Boltzmann factor
for the case with $k$ antiparallel pairs.
$x_1$ and $x_2$ have three configurations each, and one of them are displayed here
for each of them.
}
\end{center}
\end{figure}
The partition function is expressed in terms of these Boltzmann factors as
$Z_n(x_0, x_1,\cdots , x_n)$.  It is possible to apply duality transformation
to such a case, and the result is the relation
\begin{equation}
  Z_n(x_0, x_1,\cdots , x_n)=Z_n(x_0^*, x_1^*,\cdots , x_n^*).
 \label{duality-Zn}
\end{equation}
The dual Boltzmann factors $x_0^*, x_1^*,\cdots , x_n^*$ are linear functions
of the original $x_0, x_1,\cdots , x_n$.
Since the original problem has two parameters $p$ and $K$,
all the Boltzmann factors (original and dual) are functions of these two variables.

Unlike the ferromagnetic model, a naive argument using a fixed-point condition
as described above does not work in the present case because
it is in general impossible to fix all the Boltzmann factors simultaneously,
$x_0=x_0^*, x_1=x_1^*, \cdots ,x_n=x_n^*$.
The reason is that these Boltzmann factors are all functions of $p$ and $K$,
and the number of simultaneous equations $x_0(p, K)=x_0^*(p, K), x_1(p, K)=x_1^*(p, K),
\cdots , x_n(p, K)=x_n^*(p, K)$ is in general
more than sufficient to be solvable for $p$ and $K$.

We nevertheless proceed further and try the ansatz that the fixed-point condition
of the leading Boltzmann factor $x_0=x_0^*$ may lead to the correct location of
transition point if we restrict ourselves to the NL.
The results presented in Table 1 were derived this way.
We have several reasons to choose the combination of $x_0=x_0^*$ and NL
in our effort to locate the multicritical point, an important one
being that the fixed-point conditions of the other variables, $x_k=x_k^*~
(k\ne 0)$, lead to inconsistency; they violate a rigorous inequality
on the location of the multicritical point and, also, the quenched limit
of $x_k=x_k^*$ (plus NL) does not have a meaningful solution unless $k=0$.

Our conjecture that $x_0=x_0^*$ plus NL gives the exact location of the
multicritical point has been confirmed to be rigorously justified  explicitly for
$n=1, 2$ and $\infty$.  It has also  been shown by
numerical simulations that the same is very likely to be true in the
case of $n=3$.  As mentioned in the previous section, numerical evidence
for the quenched limit $n\to 0$ has also been accumulated extensively.

\section{Generalizations}

It is possible to generalize the idea described in the previous section
for the simple self-dual Ising model to a variety of other systems.

If the model is not self dual, as in the random Ising model
on the triangular and hexagonal lattices, the duality relation
(\ref{duality-Zn}) is replaced with a similar expression with
the left hand side being the partition function for the triangular
lattice and the right hand side for the hexagonal lattice, for example.
As was the case in the non-random systems, it is not possible to exactly identify
the multicritical point from such a duality relation, but we may relate the
multicritical points of the two systems.  Our conjecture for such a problem
is that the multicritical points $p_{c1}$ and $p_{c2}$ for mutually-dual
systems are related by the symmetrized generalization of $x_0=x_0^*$,
\begin{equation}
  x_0(p_{c1}, K_{c1})x_0(p_{c2}, K_{c2})=x_0^*(p_{c1}, K_{c1})x_0^*(p_{c2}, K_{c2}),
 \label{non-self-dual}
\end{equation}
where $K_{c1}$ and $K_{c2}$ are functions of $p_{c1}$ and $p_{c2}$, respectively,
through the NL condition $e^{-2K}=(1-p)/p$.
Rigorous validity of this conjecture has been verified for $n=1, 2$ and $\infty$,
and numerical simulations are in good agreement as in Table 1
for the triangular-hexagonal pair as well as for the three-dimensional problem
of the Ising spin glass and random lattice gauge system.
Equation (\ref{non-self-dual}) can also be applied to a system with anisotropy
in the probability variable $p$ on a self-dual lattice.  The $p_{c1}$
should then be understood as the critical probability for the vertical bonds
and $p_{c2}$ is for horizontal bonds on the square lattice.

Non-Ising spins can also be treated within the present theoretical framework.
It is known that the ferromagnetic Potts model is self dual on a self-dual lattice.
The random version of the Potts model, with chiral-type randomness, fits
very well to our theory using the same condition $x_0=x_0^*$ plus NL,
and the result agrees with analytical (where applicable) and numerical evidence
as indicated in Table 1.

The random chiral $Z_q$ model is another example of non-Ising system.
In the non-random ferromagnetic case, it is known that the $Z_q$ model
has two phase transitions, and therefore three thermodynamic phases, if
$q$ is larger than or equal to five\cite{JKKN}.
The random version is expected to share this feature on the NL line (and
off as well), and the two transition points are related by a
condition corresponding to (\ref{non-self-dual}).

An interesting consequence  can be derived on self-dual lattice, the square lattice
for example, that there is no finite-temperature
phase transition for the symmetric distribution $p=1/2$
if we use an identity that relates $Z_n$ and $Z_{n-1}$.
The replicated partition function is a function of $p$ and $K$ through
$x_0, x_1,\cdots , x_n$, which will here be indicated in terms of the
expression $Z_n(p, K)$.  We also write
$Z_n({\rm NL}, K)$ when $p$ is related to $K$ through the NL condition.
Then, it is possible to prove an identity $Z_n(p=1/2, K)=Z_{n-1}({\rm NL}, K)$,
up to a trivial factor.
The quenched limit of the symmetric system ($n\to 0$) is therefore equivalent
to the problem of replica number $-1$ on the NL.  This means that the transition point
of the symmetric system in the quenched limit is equal to the location of
the multicritical point of the $(-1)$-replicated system.
Thus, if we are allowed to use the replica method to the ($-1$)-replica case,
we combine the fixed-point condition
$x_0=x_0^*$ and the NL relation and apply the resulting formula to $n\to -1$
to obtain the conjectured value of the critical point for the
symmetric system. The result is that $K_c\to \infty$ or $T_c\to 0$
as $n\to 0$ for the symmetric case.  The same argument applies not only to the Ising model
but also to other systems including the random Potts and $Z_q$ models.
This is consistent with many numerical investigations and resolves some
controversies on the existence/absence problem of a finite-temperature
transition in the symmetric system on the square lattice\cite{Matsubara,Holme}.

\section{Discussions}

After briefly reviewing the present status of the theory of static properties
of spin glasses, we have presented our conjecture on the exact location of
the multicritical point in the phase diagram.
An important point to be emphasized is that all the
numerical results of independent calculations by a number of groups
can be understood in a single unified framework of our theory
that exploits duality and gauge symmetry.
If our conjecture is indeed correct, against which we have no evidence so far,
it will be the first analytical theory to systematically reveal lattice-specific properties,
{\it i.e.} the value of the critical point, of finite-dimensional systems
for which numerical methods have long been the only path to reach 
reliable conclusions.

A frequently asked question is why we restrict ourselves to the NL.
The fixed-point condition of the leading Boltzmann factor $x_0(p, K)=x_0^*(p, K)$
relates $p$ and $K$ and may give the shape of the whole part of phase boundary.
This is indeed the case for $n=1$ as well as for $n=2$ above the multicritical point.
We nevertheless restrict ourselves to the NL at this moment because
it is difficult to directly verify numerically this ansatz on the whole part of
the phase diagram for many systems with quenched randomness.
It should also be kept in mind that we have focused our attention to the point
where two completely different types of symmetries, invariance under
duality ($x_0=x_0^*$) and gauge transformation (NL condition), meet.
This implies that the multicritical point has clearly distinguished
symmetry features, which allows us to discuss this point on a different
basis from other points of phase transition.
Rigorous proof of our conjecture is awaited for.

\section*{Acknowledgements}

This work was supported by the Grant-in-Aid for Scientific Research
on Priority Area ``Statistical-Mechanical Approach to Probabilistic
Information Processing''.

\end{document}